\documentclass{article}
\usepackage{cite}
\usepackage{url}
\usepackage{hyperref}
\usepackage{graphicx}
\usepackage{subcaption}
\usepackage{PRIMEarxiv}
\usepackage{booktabs}
\usepackage{makecell}
\usepackage[utf8]{inputenc} 
\usepackage[T1]{fontenc}    
\usepackage{hyperref}       
\usepackage{url}            
\usepackage{booktabs}       
\usepackage{amsfonts}       
\usepackage{nicefrac}       
\usepackage{microtype}      
\usepackage{lipsum}
\usepackage{fancyhdr}       
\usepackage{graphicx}       
\graphicspath{{media/}}     
\usepackage{listings}
\usepackage{color}
\definecolor{dkgreen}{rgb}{0,0.6,0}
\definecolor{gray}{rgb}{0.5,0.5,0.5}
\definecolor{mauve}{rgb}{0.58,0,0.82}
\title{Achieving Tool Calling Functionality in LLMs Using Only Prompt Engineering Without Fine-Tuning
\thanks{\textit{\underline{Citation}}: 
\textbf{Authors. Title. Pages.... DOI:000000/11111.}} 
}

\author{
  Shengtao He  \\
  Hunan University \\
  Changsha, Hunan\\
  \texttt{\href{mailto:hst97@qq.com}{hst97@qq.com}} \\
}

\begin{document}
\maketitle

\begin{abstract}
Currently, the vast majority of locally deployed open-source large language models (LLMs) and some commercial model interfaces do not support stable tool calling functionality. The existing solution involves fine-tuning LLMs, which results in significant time and computational resource consumption. This paper proposes a method that enables LLMs to achieve stable tool calling capabilities using only prompt engineering and some ingenious code design. We conducted experiments on multiple LLMs that lack tool calling capabilities across various tool calling tasks, achieving a success rate of 100\%.
\end{abstract}


\section{Introduction}
Currently, the primary method for enabling large language models (LLMs) to achieve tool calling capabilities is through fine-tuning. For example, ToolLLM is a general framework for tool usage. Y. Qin et al. proposed a fine-tuning technique to enable tool calling capabilities in LLMs, providing a comprehensive API dataset \cite{qin2023toollama}. After extensive training on a large number of real API datasets, the ToolLlama model achieved stable tool calling capabilities. I. Abdelaziz et al. enabled function calling capabilities (i.e., tool calling) in LLMs through fine-grained multi-task learning \cite{abdelaziz2024granite}. Fine-tuning LLMs requires significant time and computational resources, and may even result in an LLM with lower intelligence levels than before fine-tuning. This leads to high trial-and-error costs. When a different type or more complex API structure needs to be called, the fine-tuned LLM requires additional time and computational resources for further fine-tuning to meet new demands. This is clearly detrimental to the practical application of LLMs in industry.

Prompt engineering can enable LLMs to achieve tool calling capabilities with almost no cost and high efficiency. By dynamically adjusting prompts for different application scenarios, LLMs can adapt to new tool libraries. However, the drawback of using prompt engineering is its instability. This paper proposes a method to achieve stable tool calling capabilities in LLMs using only prompt engineering.

\section{Principle}

The prompt engineering method used in this paper consists of two main parts: prompt injection and tool result feedback. Prompt injection is used to add tool information and prompts for using the tool into the system prompt. Tool result feedback involves parsing the output of the tool calling and embedding the content returned by the tool back into the LLMs.

During the prompt injection phase, the prompts used are as follows:

\begin{lstlisting}
if tools is not None:
    tools_dis = json.loads(tools)
    for tool_dis in tools_dis:
        tools_list.append(tool_dis["function"])
    tools_instructions = ""
    tools_instruction_list = []
    for tool in tools_list:
        tools_instruction_list.append(tool["name"])
        tools_instructions += (
            str(tool["name"])
            + ":"
            + "Call this tool to interact with the "
            + str(tool["name"])
            + " API. What is the "
            + str(tool["name"])
            + " API useful for? "
            + str(tool["description"])
            + ". Parameters:"
            + str(tool["parameters"])
            + "Required parameters:"
            + str(tool["parameters"]["required"])
            + "\n"
        )
TOOL_EAXMPLE = "You will receive a JSON string containing a list of callable tools. Please parse this JSON string and return a JSON object containing the tool name and tool parameters. Here is an example of the tool list:\n\n{\"tools\": [{\"name\": \"plus_one\", \"description\": \"Add one to a number\", \"parameters\": {\"type\": \"object\",\"properties\": {\"number\": {\"type\": \"string\",\"description\": \"The number that needs to be changed, for example: 1\",\"default\": \"1\",}},\"required\": [\"number\"]}},{\"name\": \"minus_one\", \"description\": \"Minus one to a number\", \"parameters\": {\"type\": \"object\",\"properties\": {\"number\": {\"type\": \"string\",\"description\": \"The number that needs to be changed, for example: 1\",\"default\": \"1\",}},\"required\": [\"number\"]}}]}\n\nBased on this tool list, generate a JSON object to call a tool. For example, if you need to add one to number 77, return:\n\n{\"tool\": \"plus_one\", \"parameters\": {\"number\": \"77\"}}\n\nPlease note that the above is just an example and does not mean that the plus_one and minus_one tools are currently available."

REUTRN_FORMAT="{\"tool\": \"tool name\", \"parameters\": {\"parameter name\": \"parameter value\"}}"

INSTRUCTION = f"""
{TOOL_EAXMPLE}
Answer the following questions as best you can. You have access to the following APIs:
{tools_instructions}
        
Use the following format:
```tool_json
{REUTRN_FORMAT}
``` 

Please choose the appropriate tool according to the user's question. If you don't need to call it, please reply directly to the user's question. When the user communicates with you in a language other than English, you need to communicate with the user in the same language.

When you have enough information from the tool results, respond directly to the user with a text message without having to call the tool again.
"""
system_prompt=INSTRUCTION
\end{lstlisting}

INSTRUCTION is the final string injected into the system prompt, which includes three parts: TOOL EXAMPLE, tools instructions, and RETURN FORMAT. TOOL EXAMPLE is used to guide the LLMs on how to understand and use the tool. When writing TOOL EXAMPLE, it is important to use trivial tools as examples, such as the tools used in this paper for incrementing and decrementing numbers, to avoid confusing the LLMs with actual usable tools. tools instructions is a list of currently available tools converted into a format readable by the LLMs. When using the LLMs in practice, tools instructions can be dynamically adjusted by inputting different tools, allowing the LLMs to know which tools are available and how to use them. RETURN FORMAT defines the format for calling the API.

During the tool result feedback phase, regular expressions are used to extract the “tool” and “parameters” from the output. For the interpreter tool, another regular expression is used to extract the code output by the LLMs, increasing the success rate of the LLMs using the interpreter tool. The code used in this paper is as follows:

\begin{lstlisting}
history.append({
    "role": "user",
    "content": user_prompt.strip()
})
response= model.create_chat_completion(
    messages = history,
    max_tokens=max_length,
    temperature=temperature,
)
response_content=response['choices'][0]['message']['content']
pattern = r'\{\s*"tool":\s*"(.*?)",\s*"parameters":\s*\{(.*?)\}\s*\}'             
while re.search(pattern, response_content, re.DOTALL)!=None:
    match=re.search(pattern, response_content, re.DOTALL)
    tool = match.group(1)
    parameters = match.group(2)
json_str = '{"tool": "' + tool + '", "parameters": {' + parameters + '}}'
parameters = json.loads('{' +parameters+ '}')
    results = dispatch_tool(tool, parameters)
    print(results)
    history.append({"role":"assistant", "content": json_str})
    history.append({"role": "observation", "content": results})
    response= model.create_chat_completion(
        messages = history,
        max_tokens=max_length,
        temperature=temperature,
    )
response_content = response.choices[0].message.content
pattern = r"```python\n(.*?)\n```"
while re.search(pattern, response, re.DOTALL) !=None:
    matches = re.search(pattern, response, re.DOTALL)
code = matches.group(1)    
results = interpreter(code)
print(results)
json_str = '{"tool": "interpreter", "parameters": '+code+'}'
    history.append({"role":"assistant", "content": json_str})
    history.append({"role": "observation", "content": results})
    response= model.create_chat_completion(
        messages = history,
        max_tokens=max_length,
        temperature=temperature,
    )
    response_content = response.choices[0].message.content
\end{lstlisting}

By identifying the dictionary of tools called by the LLM and extracting the corresponding values, these values are then passed into the appropriate tool functions. Finally, the results returned by the tools are sent back to the LLM in the role of “observation.” For some LLM interfaces that do not accept the roles of “observation,” “tool,” or “function,” the results can be returned to the “user” role instead. For example:

\begin{lstlisting}
history.append({"role": "user", "content": "Call" + tool + "The result returned by the tool is:" + results + ". Please continue to answer my previous question based on the result returned by the tool."})
\end{lstlisting}

By using the above prompt engineering method, it is possible to avoid fine-tuning and enable LLMs that originally lack tool calling capabilities to achieve stable tool calling functionality.

\section{Experimental Results}
In this study, we used the quantized versions of the current mainstream small open-source models llama3-8b, gemma2-9b, qwen2-7b, and mistral-7b from Ollama as test models \cite{ollama2024quantized}. The following tool calling tasks were tested, each with 10 different queries:\\

1. Querying real-time time in different time zones.\\
2. Querying real-time weather in different locations.\\
3. Answering recent events after performing a Google search.\\
4. Solving mathematical problems using a Python interpreter.\\
5. Searching local file information to answer questions.\\
6. Querying relevant papers on arXiv.\\
7. Searching local knowledge graphs to answer questions.\\

The tests were conducted on an NVIDIA GeForce RTX 4080, using a platform developed by the author: the open-source project ComfyUI LLM Party \cite{he2024comfyui}. This project is available on GitHub. To quickly reproduce the results of this paper, you can download the project for testing. Table 1 shows the number of successful tool calls for multiple models across various tool calling tasks using the prompt engineering method proposed in this paper. For models that do not use the prompt injection method, tool information is passed to these models via the tool interface. However, since these models do not support tool calling functionality, they cannot call these tools.

\begin{table}[h]
  \caption{Number of Successful Tool Calls Using Prompt Engineering}
  \centering
  \begin{tabular}{lccccccc}
    \toprule
    Model       & \makecell{Time Zone \\ Query} & \makecell{Weather \\ Query} & \makecell{Google \\ Search} & \makecell{Python \\ Interpreter} & \makecell{Local File \\ Search} & \makecell{ArXiv \\ Query} & \makecell{Knowledge Graph \\ Search} \\
    \midrule
    llama3-8b   & 10              & 10            & 10            & 1                  & 10                & 10          & 10                     \\
    gemma2-9b   & 10              & 10            & 10            & 10                 & 10                & 10          & 10                     \\
    qwen2-7b    & 10              & 10            & 10            & 10                 & 10                & 10          & 1                      \\
    mistral-7b  & 10              & 10            & 10            & 2                  & 10                & 10          & 1                      \\
    \bottomrule
  \end{tabular}
  \label{tab:tool-calling}
\end{table}

Table \ref{tab:tool-calling} shows that all models successfully executed the tool calling step and correctly output dictionaries that could be captured by regular expressions. However, due to limitations in code generation capabilities, the Ollama-quantized versions of the llama3-8b and mistral-7b models did not consistently output correct code in the Python interpreter task, resulting in unstable completion of computational tasks. In the knowledge graph search task, all models successfully returned relevant knowledge using the tools. However, due to limitations in logical understanding capabilities, the Ollama-quantized versions of the qwen2-7b and mistral-7b models could not consistently understand the logical relationships between multiple edges in the knowledge graph.

These experimental results demonstrate that prompt engineering can enable LLMs that originally lack tool calling capabilities to achieve tool calling functionality. However, the ability to effectively utilize the information returned by the tools to solve user problems is still limited by the LLM’s own intelligence level. Larger models, such as gemma2-9b, show significantly more stable capabilities in utilizing the results returned by the tools.

\begin{figure}[h]
  \centering
  \begin{subfigure}[b]{0.45\textwidth}
    \centering
    \includegraphics[width=\textwidth]{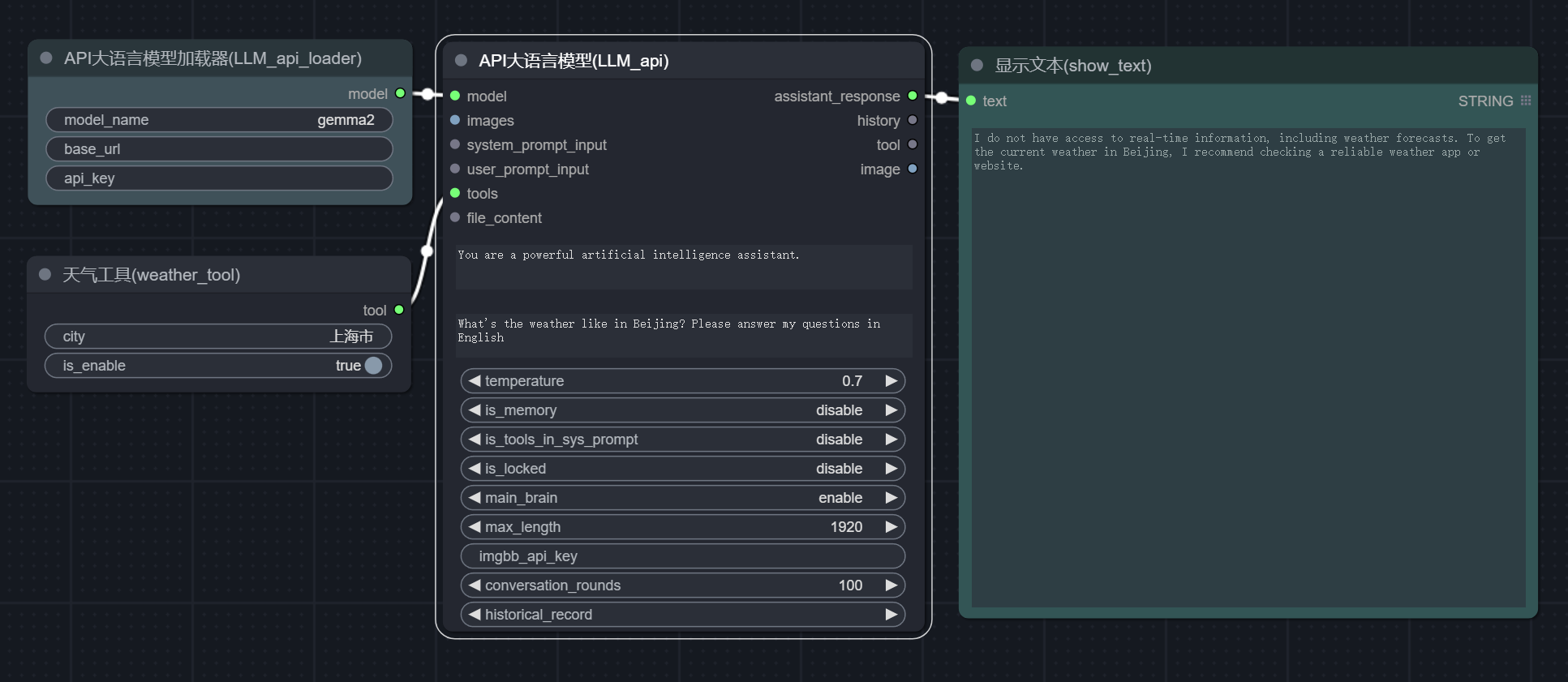}
    \caption{}
    \label{fig:sub1}
  \end{subfigure}
  \hfill
  \begin{subfigure}[b]{0.45\textwidth}
    \centering
    \includegraphics[width=\textwidth]{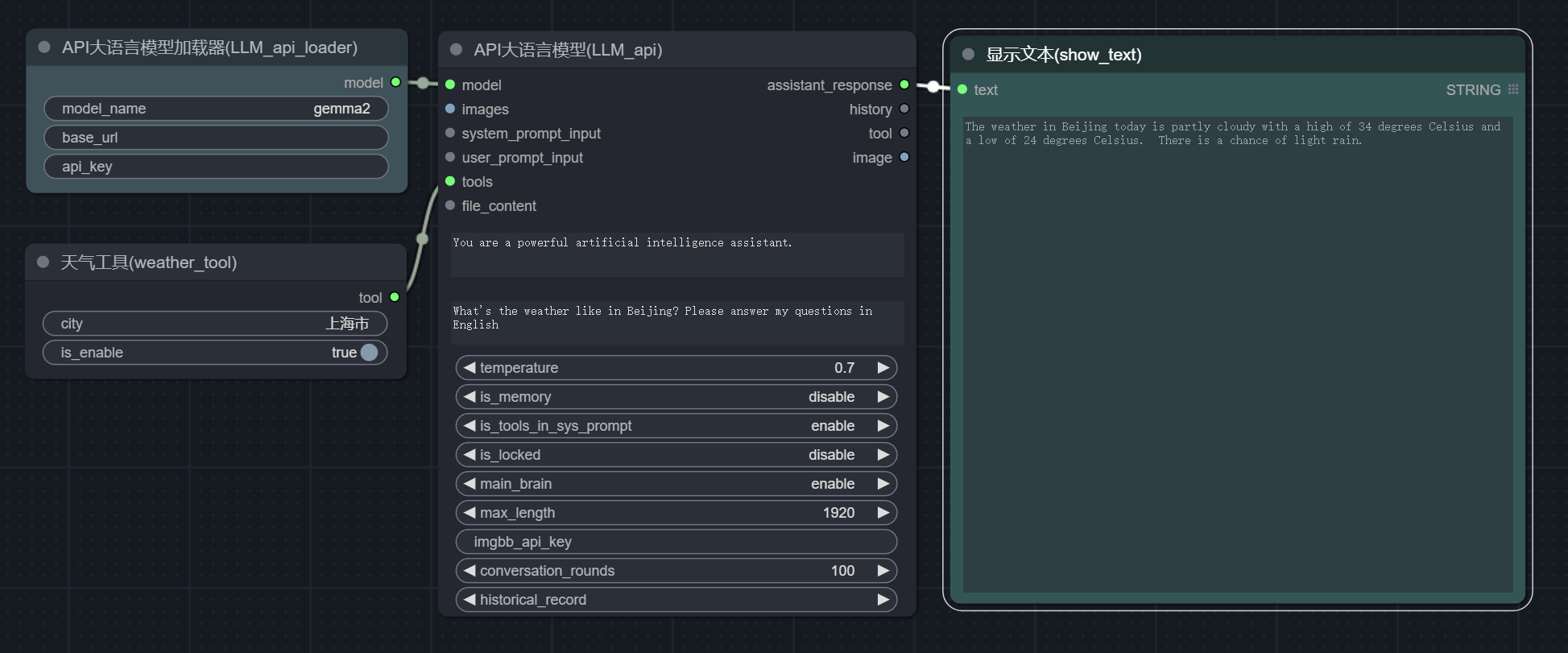}
    \caption{}
    \label{fig:sub2}
  \end{subfigure}
  \caption{(a) Output of gemma2-9b without using prompt engineering; (b) Output of gemma2-9b using prompt engineering}
  \label{fig:main}
\end{figure}

The Figure \ref{fig:main} shows the output of the gemma2-9b model when calling the weather tool. When prompt engineering is not used, the ‘is tools in sys prompt’ attribute is set to disable, and the gemma2-9b model believes it cannot obtain real-time information, thus rejecting the user’s request. When prompt engineering is used, the ‘is tools in sys prompt’ attribute is set to enable, and the gemma2-9b model provides the correct real-time weather information.

\section{Conclusion}
This study demonstrates that prompt engineering alone can enable LLMs to achieve tool calling capabilities, significantly saving time and computational resources required for fine-tuning. All models in the experiment successfully output the correct format for regular expression recognition. However, due to the intelligence level limitations of small LLMs, some models lacked the programming or logical capabilities to produce correct results for certain complex tasks. Due to the limitations of the author’s equipment, the effects of prompt engineering were not tested on larger LLMs. Researchers who have doubts about the experimental results can use the author’s open-source project on GitHub, ComfyUI LLM Party \cite{he2024comfyui}, to reproduce the work presented in this paper.

\bibliographystyle{unsrt}  
\bibliography{references}

\begin{thebibliography}{1}

\bibitem{qin2023toollama}
Y.~Qin et~al.
\newblock Toolllm: Facilitating large language models to master 16000+ real-world apis.
\newblock {\em arXiv preprint arXiv:2307.16789}, 2023.

\bibitem{abdelaziz2024granite}
I.~Abdelaziz et~al.
\newblock Granite-function calling model: Introducing function calling abilities via multi-task learning of granular tasks.
\newblock {\em arXiv preprint arXiv:2407.00121}, 2024.

\bibitem{ollama2024quantized}
Ollama.
\newblock Quantized versions of llama3-8b, gemma2-9b, qwen2-7b, and mistral-7b, 2024.
\newblock Ollama Documentation, \url{https://ollama.com/library}.

\bibitem{he2024comfyui}
S.~He.
\newblock Comfyui llm party: An open-source project for llm tool calling, 2024.
\newblock GitHub Repository, \url{https://github.com/heshengtao/comfyui_LLM_party}.

\end{thebibliography}

\end{document}